\documentclass[a4paper,11pt]{article}

\usepackage[square, numbers]{natbib}
\usepackage{amsmath}
\usepackage{amssymb}
\usepackage{arydshln}
\usepackage{yfonts}
\usepackage{amsfonts}
\usepackage{amssymb}
\usepackage{graphicx}
\usepackage[autostyle]{csquotes}
\usepackage{breqn}
\usepackage{cancel}
\usepackage[default]{fncychap}
\usepackage{epstopdf}
\usepackage[a4paper,top=2.5cm,bottom=2.5cm,left=3cm,right=3cm,headsep=10pt]{geometry}
\usepackage{fancyhdr}
\fancypagestyle{plain}{
	\fancyhf{}
	\lhead{\textit{\today}}
	\chead{\thepage}
	}
\usepackage{yfonts}
\usepackage{framed}
\usepackage[colorlinks=true, allcolors=blue]{hyperref} 

\begin{document}

\title{Residual Symmetries and BRST Cohomology of Schwarzschild in the Kerr-Schild Double Copy}

\author{\Large{Brandon Holton} \\ fgfg28@durham.ac.uk \\ Department of Mathematical Sciences \\ Durham University, UK}

\maketitle

\begin{abstract}
The Kerr-Schild (KS) double copy is celebrated for producing exact gravitational spacetimes from gauge fields, yet the preservation of symmetry content remains largely unexplored. We investigate the fate of residual symmetries in the KS double copy, focusing on the Schwarzschild solution. On the gauge theory side, we derive the residual transformations that preserve the Abelian and non-Abelian KS ansatzë, finding they both form an infinite-dimensional Lie algebra parameterized by arbitrary null functions. On the gravity side, we analyze the resulting residual diffeomorphisms of the KS Schwarzschild metric. Restricting our focus to the Killing vector class of solutions, we find that the only surviving diffeomorphisms are the finite-dimensional global isometries of Schwarzschild, reducing the residual gauge algebra to the subalgebra generated by time translations and spatial rotations. This finding reveals a fundamental structural mismatch: the infinite-dimensional algebra of the gauge side admits no simple counterpart in this constrained gravitational sector. We formalize this by showing that the BRST operator for the residual symmetry is trivialized under the Killing condition. This result serves as a crucial consistency check, validating the kinematic algebraic collapse within a quantum field theoretic framework. This paper is the first of a two-part series. In the second paper, we complete this analysis by examining the more complex proper conformal Killing vector (CKV) class of solutions and formulating a unified BRST framework to definitively test the structural obstruction.

\end{abstract}
\thispagestyle{empty}
\newpage
\pdfbookmark[0]{Contents}{contents_bookmark}
\tableofcontents
\thispagestyle{empty}
\pagenumbering{arabic}
\section{Introduction}

\subsection{Background and Motivation}

The discovery of deep structural connections between gauge theory and gravity has reshaped our understanding of field theory and spacetime. One of the most striking of these connections is the double copy, a correspondence in which gravitational field theories emerge as “squares” of gauge theories. This idea was originally motivated by the Kawai–Lewellen–Tye (KLT) relations in string theory \cite{Kawai:1986arb}, and was further refined by the Bern–Carrasco–Johansson (BCJ) color-kinematics duality \cite{Bern:2010yg, Bern:2019nnu, Bern:2019prr}. In recent years, the double copy has been extended to classical field configurations, inspiring a wide range of research programs aimed at exploring its algebraic foundations and physical implications.
\\
\\
While the earliest incarnations of the double copy were discovered in the context of string scattering amplitudes \cite{Dunbar:1994bn, Kawai:1986arb}, successful applications have since been found in mathematics \cite{Alkac:2021bav, Coll:2000rm, Easson:2023dbk}, particle physics \cite{Adamo:2020qru, Bern:2019nnu, Dunbar:1994bn, Kawai:1986arb, Monteiro:2015bna}, black hole physics \cite{Ayon-Beato:2015nvz, Gonzo:2021drq}, supersymmetry and supergravity \cite{Anastasiou:2014qba, Anastasiou:2016csv, Anastasiou:2017nsz, Cardoso:2016ngt, Cardoso:2016amd}, and quantum gravity \cite{Bern:2010ue}. Several frameworks now extend it into the classical regime. These include the self-dual \cite{Adamo:2020qru, Anastasiou:2018rdx, Campiglia:2021srh}, convolutional \cite{Godazgar:2022gfw, Luna:2020adi}, and Kerr-Schild (KS) double copy, first developed by Monteiro, O’Connell, and White \cite{Monteiro:2014cda, Monteiro:2015bna}. Each provides a distinct map between gauge fields and gravitational solutions, offering complementary insights into the structure of the correspondence.
\\
\\
The convolutional double copy is perhaps the most algebraically transparent. It constructs linearized gravitational fields by convolving pairs of Yang–Mills fields (including ghosts), preserving both linearity and BRST invariance. Within this framework, BRST symmetry plays a central role: the BRST operator $\mathcal{Q}$ consistently encodes gauge redundancies, and its cohomology identifies the physical states. Remarkably, the convolutional double copy preserves this cohomological structure, with gauge theory ghosts mapping cleanly to gravitational diffeomorphism ghosts. In this way, the convolutional formalism provides a systematic, symmetry-preserving correspondence between Yang–Mills theory and gravity \cite{Anastasiou:2018rdx, Godazgar:2022gfw, Luna:2020adi}. However, its scope is limited: because the construction is intrinsically linear, it reproduces solutions such as the Schwarzschild metric only in their linearized form. The inability to generate fully non-linear geometries motivates the search for alternative approaches.
\\
\\
The Kerr-Schild double copy, by contrast, is capable of producing exact classical solutions, having demonstrated on several accounts to provide a direct map between classical solutions in the two theories. A canonical example is the Schwarzschild–Coulomb correspondence \cite{Monteiro:2014cda, Monteiro:2015bna}. Expressed in Kerr-Schild form, the Schwarzschild metric arises from the single copy of the Abelian Coulomb potential. The structural dictionary identifies the gravitational mass $M$ with the gauge theory electric charge $Q$, the gravitational coupling $\kappa$ with the Yang–Mills coupling $g$, and the Kerr-Schild metric with its vectorial counterpart. This makes the KS formalism particularly powerful: it realizes exact black hole geometries as double copies of simple point-charge configurations.
\\
\\
What remains unclear, however, is whether the Kerr-Schild construction also preserves the underlying residual symmetries. In Yang–Mills theory, gauge transformations that preserve the Kerr-Schild ansatz form infinite-dimensional algebras. In the convolutional double copy, BRST invariance ensures that these residual symmetries lift consistently to diffeomorphisms in gravity, preserving the algebraic structure. For the Kerr-Schild double copy, by contrast, no analogous demonstration exists: while exact spacetimes such as Schwarzschild are faithfully reproduced, the status of the associated residual symmetry algebras remains unresolved.
\\
\\
Addressing this problem is the central aim of this work. On the gauge theory side, we systematically derive the full set of residual transformations preserving the KS ansatz, compute their algebras, and establish their coordinate-independence. On the gravity side, beginning with the Schwarzschild solution in Kerr-Schild form, we derive the corresponding system of PDEs for residual diffeomorphisms, solve them explicitly in the Killing sector, and analyze their algebraic structure. This comparison reveals a striking structural mismatch: while gauge theory admits infinite-dimensional residual algebras, the gravitational residual diffeomorphisms are found to reduce to the finite-dimensional global isometries of Schwarzschild, suggesting an obstruction to a symmetry-preserving map.
\newpage
\noindent
In parallel, we take a crucial first step toward a BRST formulation of the Kerr-Schild double copy, constructing a consistent ghost sector and nilpotent charge. We demonstrate that, in this sector, the residual symmetry algebra has only a trivial realization in cohomology, which serves as a quantum consistency check that validates the kinematic algebraic reduction. Together, these results provide the first systematic derivation and algebraic analysis of residual symmetries in the Kerr-Schild double copy, highlighting both the extent and the initial algebraic challenge of the correspondence at the level of symmetries.
\subsection{Outline}

The remainder of this paper is organized as follows. \textbf{Section 2} derives the residual symmetries of Abelian and non-Abelian gauge fields in the Kerr-Schild ansatz and analyzes the associated infinite-dimensional Lie algebras. \textbf{Section 3} solves the PDEs for residual diffeomorphisms in Schwarzschild spacetime, identifies the surviving Killing sector solutions, and compares their finite-dimensional algebra with that of the gauge theory, highlighting the structural mismatch. This section then develops the BRST formulation, which serves as a crucial quantum consistency check, establishing the cohomological validation of the Killing vectors. \textbf{Section 4} summarizes the results, highlights the initial algebraic challenge and limitations of the KS double copy at the symmetry level, and discusses future directions, including the systematic analysis of the conformal Killing vector class, extensions to the Kerr spacetime, and other double copy frameworks.
\subsection{Conventions}

Throughout both papers in this series, we adopt the mostly-plus convention $( -, +, +, + )$ in this paper, and, unless explicitly noted, the background is taken to be flat Minkowski space $\eta_{\mu\nu}$. In Cartesian coordinates $(t,x,y,z)$, the background metric is $\eta_{\mu \nu} = \text{diag} (-1, 1, 1, 1).$ In spherical coordinates $(t,r,\vartheta,\varphi)$, it is $\eta_{\mu \nu} = \text{diag} (-1, 1, r^2, r^2 \sin^2 \vartheta)$. Because our analysis centers on Schwarzschild geometry, spherical coordinates are the default choice, and it should be assumed that we are working in spherical coordinates unless stated otherwise.
\\
\\
Additionally, for each parametrization of the Kerr-Schild vector $k^\mu$, we take all components of $k^\mu$ to be positive. The corresponding (co)vector $k_\mu$ then carries a relative minus sign in the time component. In Cartesian coordinates, we take the convention of Monteiro, O'Connell, and White \cite{Monteiro:2014cda, Monteiro:2015bna} for $k^\mu$:

        \begin{equation}
         \label{GG} \begin{matrix}
                k^\mu = \begin{pmatrix}
                1, \frac{x^i}{r}
            \end{pmatrix} & , & k_\mu = \begin{pmatrix}
                -1, \frac{x^i}{r}
            \end{pmatrix}
            \end{matrix}
        \end{equation}
\\
for $i = 1, 2, 3$ and $x^i x_i := r^2 = x^2 + y^2 + z^2$. In spherical coordinates,

        \begin{equation}
       \label{HAPPY}  \begin{matrix}
                k^\mu = \begin{pmatrix}
                1,~1,~0,~0
            \end{pmatrix} & , & k_\mu = \begin{pmatrix}
                -1,~1,~0,~0
            \end{pmatrix}
            \end{matrix}.
        \end{equation}
\\
Naturally, we adopt the spherical form of $k^\mu$ throughout, unless stated otherwise. To our knowledge, this parameterization first appeared in Gonzo and Shi \cite{Gonzo:2021drq}, who used it to analyze particle geodesics around Kerr black holes.
\section{Yang-Mills Symmetries in Schwarzschild Spacetime}

In this section, we examine the residual symmetries of gauge theories that admit a Kerr-Schild-type ansatz, starting with the Abelian case. We identify the class of gauge transformations that preserve the functional form of the gauge potential $A_\mu$ and show how these symmetries can be characterized using the method of characteristics in spherical coordinates. These transformations form a symmetry algebra, which we compute explicitly, along with the algebra induced on the scalar field $\Phi$. After establishing the Abelian case, we extend the analysis to non-Abelian gauge theory, where self-interactions modify the symmetry structure and complicate algebraic closure. This section therefore establishes how gauge theoretic residual symmetries behave under the Kerr-Schild constraints, setting the stage for their gravitational counterparts in the double copy framework.
\subsection{The Abelian Case}

We begin by analyzing residual gauge symmetries in the Abelian theory, where the structure is simplest. Our goal is to identify the class of gauge transformations that preserve the Kerr-Schild ansatz for the gauge field, which we will introduce shortly. This reduces the problem to finding functions $\lambda(x)$ such that $\delta_\lambda A_\mu = \partial_\mu \lambda(x)$ respects the chosen field structure. Solving this in the Abelian case establishes a clear baseline and introduces the methodology we will generalize to the non-Abelian setting.
\subsubsection{Residual Symmetries in Spherical Coordinates}

Consider the Kerr-Schild ansatz for the gauge field $A_\mu$ in Schwarzschild coordinates:

\begin{equation}
   \label{A} A_\mu := \Phi(x) k_\mu,
\end{equation}
\\
where $k_\mu$ is given by \eqref{HAPPY}. Under a local gauge transformation with smooth parameter $\lambda(x)$, the field transforms as 

\begin{equation}
   \label{B} A_\mu \rightarrow A_\mu' = A_\mu + \partial_\mu \lambda(x).
\end{equation}
\\
We require that $A_\mu'$ preserve the Kerr-Schild form \eqref{A}, i.e., there exists a scalar field $\Phi'(x)$ such that 

\begin{equation}
   \label{C} A_\mu' := \Phi'(x) k_\mu.
\end{equation}
\\
For an infinitesimal perturbation $\delta_\lambda$, the transformed field is

\begin{equation}
    A_\mu' = A_\mu + \delta_\lambda A_\mu = [\Phi(x) + \delta_\lambda \Phi(x)] k_\mu.
\end{equation}
\\
Comparing with \eqref{B} gives

\begin{equation}
  \label{E}  \partial_\mu \lambda(x) = \delta_\lambda \Phi(x) k_\mu.
\end{equation}
\newpage
\noindent
Thus, only gauge transformations for which $\partial_\mu \lambda(x)$ is proportional to $k_\mu$ preserve the Kerr-Schild form. To solve for $\lambda(x)$, we exploit the null condition $k^\mu k_\mu = 0$. Projecting \eqref{E} along $k^\mu$ yields a homogeneous PDE:

\begin{equation}
    k^\mu \partial_\mu \lambda(x) = 0.
\end{equation}
\\
With $k^t = k^r = 1$ and $k^\vartheta = k^\varphi = 0$ in spherical coordinates, this reduces to

\begin{equation}
   [\partial_t + \partial_r] \lambda(x) = 0.
\end{equation}
\\
This can be solved via the method of characteristics. Define curves $s \mapsto (t(s), r(s))$ such that along these curves $d\lambda / ds = 0$. Choosing tangent vectors aligned with the PDE coefficients,

\begin{equation}
        \frac{dt}{ds} = 1 ~~~~~,~~~~~ \frac{dr}{ds} = 1,
\end{equation}
\\
we find

\begin{equation}
    \frac{dt}{dr} = 1 \implies t - r = \text{constant}.
\end{equation}
\\
Along these outgoing null curves, $\lambda$ is constant. Hence, the general solution for the residual gauge parameter is

\begin{equation}
   \label{12} \lambda(t,r) = f(t-r),
\end{equation}
\\
where $f$ is an arbitrary smooth function. The residual gauge freedom is therefore “frozen” along outgoing null rays, propagating only in the retarded time $u = t - r$. Plugging \eqref{12} into \eqref{B}, the nonzero transformed field components become

\begin{equation}
    A_t' = \Phi(x) k_t + \partial_t f(t-r) = [\Phi(x) - f_{,u}(u)] k_t ~~~~~,~~~~~A_r' = \Phi(x) k_r + \partial_r f(t-r) = [\Phi(x) - f_{,u}(u)] k_r,
\end{equation}
\\
where $f_{,u} = df/du$ and $u = t-r$. With $A_\vartheta' = A_\varphi' = 0$, the transformed field can be written compactly as

\begin{equation}
    A_\mu' = [\Phi(x) - f_{,u}(u)] k_\mu
\end{equation}
\\
so the transformed scalar field is

\begin{equation}
    \boxed{\Phi'(x) = \Phi(x) - f_{,u}(u) ~~~~~,~~~~~ \delta_\lambda \Phi(x) = - f_{,u}(u).}
\end{equation}
\\
This characterizes the Abelian residual gauge symmetry that preserves the Kerr-Schild form, showing explicitly how the gauge parameter $\lambda(x)$ is constrained along outgoing null rays. This result provides a baseline for the analysis we will extend to the non-Abelian theory in Section 2.2.
\subsubsection{Algebra Generated by Residual Symmetries}

We have established that an infinitesimal gauge transformation of the form

\begin{equation}
    \delta_\lambda A_\mu = \partial_\mu \lambda(x)
\end{equation}
\\
preserves the Kerr-Schild ansatz \eqref{A} only if

\begin{equation}
    \label{I} \partial_\mu \lambda(x) = \delta_\lambda \Phi(x) k_\mu,
\end{equation}
\\
with the additional constraint $k^\mu \partial_\mu \lambda(x) = 0$. This implies that $\lambda(x)$ must be constant along the integral curves of $k^\mu$, which correspond to outgoing radial null geodesics. In terms of the null coordinate $u = t-r$, the general solution is therefore

\begin{equation}
       \label{J} \lambda(x) = f(u) ~~~~~, ~~~~~ f(u) \in C^\infty(\mathbb{R}).
\end{equation}
\\
The resulting transformation acts as

\begin{equation}
    \delta_\lambda A_\mu = \partial_\mu \lambda(x) = f_{,u}(u) k_\mu
\end{equation}
\\
so that the gauge field transforms as

\begin{equation}
    A_\mu \rightarrow A_\mu' = A_\mu + \delta_\lambda A_\mu = [\Phi(x) - f_{,u}(u)] k_\mu = \Phi'(x) k_\mu,
\end{equation}
\\
which manifestly preserves the Kerr-Schild structure.
\\
\\
We now determine the algebra of these transformations. Let $\mathfrak{g}_{\text{res}}$ denote the space of residual gauge transformations $\delta\lambda$ with $\lambda(x) = f(u)$. Each such transformation corresponds uniquely to a smooth function $f(u) \in C^\infty(\mathbb{R})$, so there is a natural identification

\begin{equation}
\Psi : \mathfrak{g}_{\text{res}} \rightarrow C^\infty(\mathbb{R})  ~~~~~,~~~~~ \delta_\lambda \mapsto f(u).
\end{equation}
\\
This map is a vector space isomorphism: it is linear, injective (only the trivial transformation maps to $f=0$), and surjective (every $f(u)$ defines a valid residual transformation).
\\
\\
Since the gauge theory is Abelian, the commutator of two transformations vanishes:

\begin{equation}
    [\delta_{\lambda_1}, \delta_{\lambda_2}] A_\mu = 0  \Leftrightarrow [f_1, f_2] = f_1 f_2 - f_2 f_1 = 0.
\end{equation}
\\
Hence, $\Psi$ also preserves the Lie algebra structure. We conclude that the residual symmetries of the Abelian Kerr-Schild ansatz form an infinite-dimensional Abelian Lie algebra, 

\begin{equation}
    \mathfrak{g}_{\text{res}} \cong C^\infty(\mathbb{R}),
\end{equation}
\\
the additive Lie algebra of smooth functions on the null coordinate $u = t-r$.
\subsubsection{Algebra Induced over the Field $\Phi(x)$}

Before moving to the non-Abelian case, it is useful to examine how the residual symmetries act on the scalar profile $\Phi(x)$. The key point is that the Abelian residual gauge transformations correspond to functions $\lambda(x) = f(u)$ along outgoing null rays, with $u = t-r$. Physically, this means that the freedom in $\lambda$ is “frozen” along the direction of light-like propagation: any shift of $\Phi$ occurs only along the outgoing null congruence defined by $k^\mu$.
\\
\\
The infinitesimal action of a residual gauge transformation on $\Phi(x)$ is

\begin{equation}
    \delta_f \Phi(u) = - f'(u),
\end{equation}
\\
where the minus sign arises from the orientation of $k^\mu$. This describes an additive shift along outgoing null rays, so that the space of scalar field profiles naturally carries a representation of the residual gauge algebra. In other words, each smooth function $f(u)$ generates a linear operator on the space of $\Phi$ configurations, shifting the field locally along the null coordinate.
\\
\\
To understand the algebraic structure induced on $\Phi$, consider two transformations $\delta_f$ and $\delta_g$:

\begin{equation}
    [\delta_f, \delta_g] \Phi(u) = \delta_f(\delta_g \Phi) - \delta_g(\delta_f \Phi) = 0.
\end{equation}
\\
The vanishing commutator reflects that these shifts act independently along null rays: applying one transformation does not interfere with the other. Therefore, the induced algebra on the scalar field is Abelian, just like the underlying gauge algebra.
\\
\\
Not all gauge functions produce a nontrivial effect on $\Phi$. Constant functions $f(u) = c$ generate $\delta_f \Phi = 0$, leaving the field unchanged. Removing these trivial transformations gives the physically meaningful algebra:

\begin{equation}
    \boxed{\mathfrak{g}_{\text{res}} \cong C^\infty(\mathbb{R}) / \mathbb{R},}
\end{equation}
\\
which captures precisely the residual gauge freedom that manifests in $\Phi$ along outgoing null rays. In summary, the Abelian residual gauge transformations act as local shifts of the scalar field along null rays, and these shifts form an infinite-dimensional Abelian algebra modulo constants. This provides a clear, physically intuitive benchmark for understanding the more complicated non-Abelian and gravitational cases that follow.
\subsection{The Non-Abelian Case}

In the non-Abelian theory, the analysis of residual symmetries is enriched by the presence of self-interactions, which deform the structure relative to the Abelian case. The gauge transformations no longer commute, and the scalar profile $\Phi^a(x)$ transforms in the adjoint representation, introducing structure constants into the symmetry algebra. Our aim here is to generalize the Abelian construction, identifying the class of residual transformations that preserve the Kerr-Schild form and understanding the algebra they induce.
\subsubsection{Residual Symmetries in Spherical Coordinates}
Consider the Kerr-Schild ansatz for the non-Abelian gauge field,
$A_\mu^a$:

\begin{equation}
   \label{NA-KS} A_\mu^a := \Phi^a(x) k_\mu,
\end{equation}
\\
where $\Phi^a(x)$ is adjoint-valued with $a = 1,...,N^2 -1$. A finite gauge transformation acts as

\begin{equation}
      \label{NA-gauge} A_\mu^a \rightarrow A_{\mu}^{'a} = A_{\mu}^a + \partial_\mu \Lambda^a (x) + g f^{abc} A_\mu^b \Lambda^c(x),
\end{equation}
\\
with Yang-Mills coupling $g$ and gauge parameters $\Lambda^a(x)$. To preserve the Kerr-Schild structure, the transformed field must again take the form

\begin{equation}
    \label{LX} A_{\mu}^{'a} = \Phi^{'a}(x) k_\mu.
\end{equation}
\\
Writing the infinitesimal transformation as

\begin{equation}
    \label{Q} A_{\mu}^a \rightarrow  A_{\mu}^{'a} = A_{\mu}^a + \delta_\Lambda A_{\mu}^a = [\Phi^a(x) + \delta_\Lambda \Phi^a(x)] k_\mu
\end{equation}
\\
and comparing with \eqref{NA-gauge}, we obtain the preservation condition:

\begin{equation}
\delta_\Lambda \Phi^a(x) k_\mu = \partial_\mu \Lambda^a(x) + g f^{abc} \Phi^b(x) k_\mu \Lambda^c(x).
\end{equation}
\\
Projecting along $k^\mu$ and invoking the null condition $k^\mu k_\mu = 0$ eliminates the self-interaction term:

\begin{equation}
    \label{O} k^\mu \partial_\mu \Lambda^a(x) = 0.
\end{equation}
\\
Thus, as in the Abelian case, the gauge parameters must be constant along the integral curves of $k^\mu$. In spherical coordinates this gives the general solution

\begin{equation}
    \Lambda^a(t,r) = f^a(t-r) = f^a(u),
\end{equation}
\\
where $f^a(u)$ are arbitrary smooth functions. The space of residual symmetries is therefore infinite-dimensional, spanned by $N^2-1$ independent functional directions — one for each adjoint index.
\\
\\
Substituting this solution into \eqref{NA-gauge} yields the transformed field,

\begin{equation}
    A_{\mu}^{'a} = \Phi^{a}(x) k_\mu + \partial_\mu f^a(t-r) + g f^{abc} \Phi^b k_\mu f^c(t-r).
\end{equation}
\\
so that the scalar transforms as

\begin{equation}
   \boxed{\delta_\Lambda \Phi^a(x) = - f^{a}_{,u}(t-r) + g f^{abc} \Phi^b(x) f^c(t-r),}
\end{equation}
\\
This result generalizes the Abelian shift symmetry. The first term reproduces the familiar local shift along null rays, now indexed by $a$, while the second term introduces a nontrivial adjoint action proportional to the structure constants. Physically, this reflects how different color components of the field couple to one another as they propagate along the outgoing null congruence. The residual symmetry remains infinite-dimensional, but now carries the full imprint of the non-Abelian Lie algebra.
\subsubsection{Algebra Generated by Residual Symmetries}

Having established the form of the residual gauge transformations in the non-Abelian case,

\begin{equation}
\delta_\Lambda \Phi^a(x) = - f_{,u}^a(u) + g f^{abc} \Phi^b(x) f^c(u), \qquad f^a(u) \in C^\infty(\mathbb{R}),
\end{equation}
\\
we now examine the algebra they generate. Let $\mathfrak{g}_{\text{res}}$ denote the set of infinitesimal transformations $\delta_\Lambda$ with $\Lambda^a(x) = f^a(u)$. Define a map

\begin{equation}
    \Psi : \mathfrak{g}_{\text{res}} \rightarrow \mathfrak{g} \otimes C^\infty(\mathbb{R}) ~~~~~,~~~~~ \delta_\Lambda \mapsto f^a(u) T^a,
\end{equation}
\\
where $\{T^a\}$ are the generators of the Lie algebra $\mathfrak{g}$. 
\\
\\
This map is:

\begin{itemize}
    \item \textbf{Linear}: for $\delta_{\Lambda_1}, \delta_{\Lambda_2} \in \mathfrak{g}_{\text{res}}$ and $\alpha,\beta \in \mathbb{R}$,

\begin{equation}
    \Psi(\alpha \delta_{\Lambda_1} + \beta \delta_{\Lambda_2}) = \alpha f_1^a(u) T^a + \beta f_2^a(u) T^a = \alpha \Psi(\delta_{\Lambda_1}) + \beta \Psi(\delta_{\Lambda_2}).
\end{equation}

    \item \textbf{Injective}: if $\Psi(\delta_\Lambda) = 0$, then  $f^a(u) = 0$ for all $u$, implying $\Lambda^a(x) = 0$, so $\delta_\Lambda = 0$.
    
    \item \textbf{Surjective}: for any $f^a(u) \in C^\infty(\mathbb{R})$, there exists a $\delta_\Lambda \in \mathfrak{g}_{\text{res}}$ such that $\Lambda^a(x) = f^a(u)$.
\end{itemize}
\noindent
Hence, $\Psi$ is a vector space isomorphism, i.e., $\mathfrak{g}_{\text{res}} \cong \mathfrak{g} \otimes C^\infty(\mathbb{R})$. The Lie bracket is inherited pointwise from $\mathfrak{g}$

\begin{equation}
    [f^a T^a, g^b T^b] (u) := f^a g^b [T^a, T^b](u) = f^{abc} f^a(u) g^b(u) T^c,
\end{equation}
\\
where $[T^a, T^b] := f^{abc} T^c$. This bracket is bilinear, antisymmetric, and satisfies the Jacobi identity, as required.
\\
\\
Consequently, the non-Abelian residual gauge algebra forms a current algebra along outgoing null rays:

\begin{equation}
    \boxed{\mathfrak{g}_{\text{res}} \cong \mathfrak{g} \otimes C^\infty(\mathbb{R}).}
\end{equation}
\newpage
\noindent
For $\mathfrak{g} = \mathfrak{su}(N)$, this is the algebra of smooth $\mathfrak{su}(N)-$valued functions along $u = t - r$, reflecting the infinite-dimensional functional freedom in the residual gauge transformations while preserving the Kerr-Schild structure.
\subsubsection{Algebra Induced over the Field $\Phi^a(x)$}

We now examine the algebra induced on the scalar profile $\Phi^a(x)$ by the residual gauge symmetries that preserve the Kerr-Schild ansatz. While the residual transformations act on the gauge field $A_\mu^a$ via first-order differential operators, their action on $\Phi^a(x)$ encodes the physically relevant gauge-invariant information.
\\
\\
Define the infinitesimal action of a residual gauge transformation as

\begin{equation}
    \delta_f : \Phi^a(u) \mapsto - f_{,u}^a(u) + g f^{abc} \Phi^b(u) f^c(u).
\end{equation}
\\
Let $\mathcal{F}$ denote the space of admissible scalar fields $\Phi^a(u)$. The residual gauge transformations then define a linear map $\delta_f : \mathcal{F} \rightarrow \mathcal{F}$, yielding a representation of the gauge algebra $\mathfrak{g}$ on $\mathcal{F}$.
\\
\\
Since the $\delta_f$ are linear maps on $\mathcal{F}$, we may study the Lie algebra they generate via the commutator. Let $\delta_1$ and $\delta_2$ be two such transformations, defined by parameter functions $f^a(u),~h^a(u) \in C^\infty(\mathbb{R})$. Computing the commutator

\begin{equation}
    [\delta_f, \delta_h] := \delta_f \circ \delta_h - \delta_h \circ \delta_f
\end{equation}
\\
allows us to determine how these residual gauge transformations close under composition and reveals the field-dependent structure of the induced algebra. Acting on $\Phi^a(u)$, the commutator of two residual gauge transformations is defined by

\begin{equation}
     [\delta_f, \delta_h] \Phi^a(u) := \delta_f \begin{pmatrix}
         \delta_h \Phi^a(u) \end{pmatrix} - \delta_h \begin{pmatrix}
         \delta_f \Phi^a(u) \end{pmatrix}.
\end{equation}
\\
This expression measures the non-commutativity of the transformations and will reveal the nontrivial structure of the induced algebra on $\Phi^a(u)$, in contrast with the Abelian case where the commutator vanishes.
\\
\\
A straightforward but careful computation yields:

\begin{equation}
\begin{split}
        \label{OM} [\delta_f, \delta_h] \Phi^a(u) &= - g f^{abc} \begin{pmatrix}
            f_{,u}^b(u) h^c(u) - f^c(u) h_{,u}^b(u) 
        \end{pmatrix} \\ &+ g^2 f^{abc} f^{bde} \Phi^d(u) \begin{pmatrix}
            f^e(u) h^c(u) - h^e(u) f^c(u)
        \end{pmatrix},
\end{split}
\end{equation}
\\
where the first term arises from derivatives along the null direction, and the second is field-dependent, reflecting non-Abelian self-interactions. Using the antisymmetry of the structure constants and the Jacobi identity

\begin{equation}
   \label{ROX} f^{abe} f^{bcd} + f^{dab} f^{cbe} + f^{abc} f^{bde} = 0,
\end{equation}
\newpage
\noindent
the expression simplifies to

\begin{equation}
     [\delta_f, \delta_h] \Phi^a(u) = - g f^{abc} \partial_u (f^b(u) h^c(u)) + g^2 f^{abc} f^{bde} \Phi^c(u) f^d(u) h^e(u),
\end{equation}
\\
where we've taken the Leibniz rule:

\begin{equation}
    f_{,u}^b(u) h^c(u) + f^b(u) h_{,u}^c(u) = \partial_u (f^b(u) h^c(u)).
\end{equation}
\\
Define the pointwise Lie bracket on the gauge parameters:

\begin{equation}
    [f,h]^a(u) := g f^{abc} f^b(u) h^c(u).
\end{equation}
\\
and the induced transformation

\begin{equation}
    \delta_{[f,h]} := -\partial_u \begin{pmatrix} [f,h]^a \end{pmatrix} + g f^{abc} [f,h]^b \Phi^c
\end{equation}
\\
It follows immediately that

\begin{equation}
  [\delta_f, \delta_h] \Phi^a(u) = \delta_{[f,h]} \Phi^a \implies [\delta_f, \delta_h] = \delta_{[f,h]},
\end{equation}
\\
showing closure under the commutator. Consequently, the residual transformations on \(\Phi^a(u)\) form an infinite-dimensional Lie algebra, isomorphic to the current algebra

\begin{equation}
      \boxed{\mathfrak{g}_{\text{res}} \cong \mathfrak{g} \otimes C^\infty(\mathbb{R}).}
\end{equation}
\\
Although \(\mathfrak{g}_{\text{res}}\) acts linearly on the gauge parameters \(f^a(u)\), its induced action on \(\Phi^a(u)\) is \textit{nonlinear} in both gauge parameters and coupling \(g\), a direct consequence of non-Abelian self-interactions along null rays.
\\
\\
In summary:

\begin{itemize}
    \item In the Abelian case, residual transformations are parametrized by arbitrary smooth functions of $u$, forming the infinite-dimensional Abelian algebra $C^\infty(\mathbb{R})$. The induced algebra on $\Phi(x)$ reduces to $C^\infty(\mathbb{R})/\mathbb{R}$, reflecting the physical irrelevance of constant shifts.
    \item In the non-Abelian case, the structure constants introduce nonlinearity but preserve the essential null dependence. The induced algebra on $\Phi^a(u)$ is a nonlinear current algebra $\mathfrak{g} \otimes C^\infty(\mathbb{R})$, showing that the Kerr-Schild ansatz supports an infinite-dimensional symmetry structure even with self-interactions.
\end{itemize}
\noindent
This sets the stage for the gravitational analysis, where (as we will show), residual diffeomorphisms preserving the Kerr-Schild form reduce to a finite-dimensional algebra.
\section{Gravitational Symmetries in Schwarzschild Spacetime}

We now turn to gravity and study residual diffeomorphisms that preserve the Kerr-Schild form of the Schwarzschild metric. In gauge theory, the residual symmetries associated with the Kerr-Schild ansatz form infinite-dimensional algebras, but it is not clear \textit{a priori} whether a similar richness exists in gravity. To address this, we consider infinitesimal coordinate transformations generated by vector fields $\xi^\mu$ and impose the condition that the Kerr-Schild structure of the metric is maintained. In spherical coordinates, the resulting system of partial differential equations naturally decomposes into angular, radial–temporal, and mixed components. The angular subsystem admits both Killing vectors and conformal Killing vectors of the two-sphere; for clarity, we focus on the Killing sector, which captures the essential algebraic structure and leads to a tractable system.
\\
\\
Solving the Killing sector equations shows that the residual diffeomorphisms reduce to the global isometries of Schwarzschild: time translations and spatial rotations. Unlike in gauge theory, no infinite-dimensional enhancement arises in this context: the algebra of residual diffeomorphisms is finite-dimensional. This suggests an explicit algebraic mismatch between the gravitational and gauge theory sides of the Kerr-Schild double copy, highlighting a fundamental limitation of symmetry preservation in this formalism.
\subsection{Diffeomorphisms and the Lie Derivative of the Kerr-Schild Metric}

We begin by recalling the Kerr-Schild (KS) ansatz, which expresses a spacetime metric as a deformation of a background metric $\eta_{\mu\nu}$ via a scalar field $\varphi$ and a null vector $k^\mu$:

\begin{equation}
   \label{XX} g_{\mu \nu} := \eta_{\mu \nu} + \varphi k_\mu k_\nu.
\end{equation}
\\
The vector $k^\mu$ is null with respect to both the background and full metric, $k^\mu k_\mu = 0$, and geodesic with respect to the full connection, $k^\mu \nabla^{(g)}_\mu k^\nu = 0$. This guarantees an affine parametrization and implies $k^\mu$ is also geodesic with respect to the background connection. Importantly, the KS ansatz represents an exact metric; for example, Schwarzschild can be written in KS form with

\begin{equation}
    \begin{matrix}
        \varphi := \frac{2GM}{r} & , & k^\mu := (1, 1, 0, 0).
    \end{matrix}
\end{equation}
\\
in spherical coordinates $(t,r,\vartheta,\varphi)$.
\\
\\
To define residual diffeomorphisms, consider an infinitesimal transformation generated by $\xi^\mu$, under which the metric transforms as $g_{\mu\nu} \mapsto g_{\mu\nu} + \delta_\xi g_{\mu\nu} = g_{\mu\nu} + (\mathcal{L}_\xi g)_{\mu\nu}$. Preserving the KS structure requires that this change can be written

\begin{equation}
    g_{\mu \nu}' \stackrel{!}{=} \eta_{\mu \nu} + (\varphi + \delta_\xi \varphi) k_\mu k_\nu
\end{equation}
\\
so that the effect of the transformation is absorbed into a redefinition of the scalar field, $\varphi \mapsto \varphi + \delta_\xi \varphi$. Equivalently, the Lie derivative of the metric must satisfy

\begin{equation}
\label{YY}
\boxed{(\mathcal{L}_\xi g)_{\mu\nu} \stackrel{!}{=} \alpha(x) k_\mu k_\nu}
\end{equation}
\\
for some smooth function $\alpha(x)$. Vector fields $\xi^\mu$ satisfying this condition are called residual diffeomorphisms, in analogy with the residual gauge symmetries discussed in Section~2.
\\
\\
In components, the Lie derivative reads

\begin{equation}
    \label{QUACK} (\mathcal{L}_\xi g)_{\mu \nu} := \xi^\rho \partial_\rho g_{\mu \nu} + 2 \partial_{(\mu} \xi^\rho g_{\nu) \rho},
\end{equation}
\\
where $2 \partial_{(\mu} \xi^\rho g_{\nu) \rho } = \partial_\mu \xi^\rho g_{\rho \nu} + \partial_\nu \xi^\rho g_{\mu \rho}$. Substituting \eqref{XX}, we obtain a natural decomposition:

\begin{equation}
    \label{ZZ} 
        (\mathcal{L}_\xi g)_{\mu \nu} := (\mathcal{L}_\xi \eta)_{\mu \nu} + \mathcal{L}_\xi (\varphi k_\mu k_\nu),
\end{equation}
\\
with

\begin{equation}
    \label{AAA} \begin{matrix}
        (\mathcal{L}_\xi \eta)_{\mu \nu} := \underbrace{\xi^\rho \partial_\rho \eta_{\mu \nu}}_{(1)} + \underbrace{2 \partial_{(\mu} \xi^\rho \eta_{\nu)\rho}}_{(2)} & , & \mathcal{L}_\xi (\varphi k_\mu k_\nu) := \underbrace{(\mathcal{L}_\xi \varphi) k_\mu k_\nu}_{(3)} + \underbrace{2 \varphi k_{(\mu} \mathcal{L}_\xi k_{\nu)}}_{(4)}.
    \end{matrix}
\end{equation}
\\
Here, $(\mathcal{L}_\xi \eta)_{\mu\nu}$ captures the background variation, the second term describes the flow of the scalar field, and the third term captures the variation of the null vector. In spherical coordinates, $(\mathcal{L}_\xi \eta)_{\mu\nu}$ does not vanish due to explicit coordinate dependence, while $(\mathcal{L}_\xi k)_\mu$ simplifies but is generally nonzero.
\\
\\
The residual diffeomorphism condition $(\mathcal{L}_\xi g)_{\mu \nu} \propto k_\mu k_\nu$ thus constrains $\xi^\mu$ in terms of the background, the null vector, and the scalar profile. In the next section, we solve the resulting PDEs. The angular subsystem admits both Killing and conformal Killing vectors of the two-sphere; for clarity, we analyze the Killing sector first, which yields a tractable system with constant coefficients. The conformal sector introduces gradient-type solutions and will be treated separately.
\subsection{Deriving the General Class of Residual Diffeomorphisms}

To determine the diffeomorphisms that preserve the Kerr-Schild form of the Schwarzschild metric, we translate condition \eqref{YY} into a system of partial differential equations. This system is highly constrained and naturally decomposes into three subsets: (i) angular equations, (ii) radial–temporal equations, and (iii) mixed equations coupling $(t,r)$ to $(\vartheta,\phi)$. Our strategy is to analyze each subsystem sequentially: first solving the angular sector, then the radial–temporal sector, and finally imposing the mixed equations as consistency conditions. The angular subsystem formally admits both Killing and conformal Killing vectors of the round 2–sphere; for clarity, we focus on the Killing sector, which yields a closed, tractable system. Within this sector, the mixed equations enforce constant coefficients, allowing the angular Killing vectors to be consistently lifted to global isometries of the full Schwarzschild metric.
\newpage
\noindent
Before proceeding, it is useful to recall the relevant Lie derivatives. The action of a vector field $\xi^\mu$ on the scalar field $\varphi$ is simply

\begin{equation}
   \label{FFFF} \mathcal{L}_\xi \varphi := \xi(\varphi) = \xi^\mu \partial_\mu \varphi
\end{equation}
\\
while its action on the null (co)vector $k_\mu$ is

\begin{equation}
     (\mathcal{L}_\xi k)_\mu := \xi^\rho \partial_\rho k_\mu + k_\rho \partial_\mu \xi^\rho.
\end{equation}
\\
Geometrically, the first term describes the intrinsic transport of $k_\mu$ along the flow generated by $\xi^\mu$, while the second term captures the deformation of the coordinate grid, including stretching and shearing effects. In spherical coordinates, where $k^\mu = (1,1,0,0)$ is constant, we have $\partial_\mu k_\nu = 0$, so the Lie derivative simplifies to
\\
\\
Equation \eqref{FFFF} illustrates the utility in adopting spherical coordinates: since the null vector $k^\mu$ is constant, it satisfies $\partial_\mu k_\nu = 0$ for all indices $\mu, \nu$. As a result, the Lie derivative reduces to

\begin{equation}
    \label{QQQQ} (\mathcal{L}_\xi k)_\mu := k_\rho \partial_\mu \xi^\rho.
\end{equation}
\\
Expanding the full Lie derivative \eqref{ZZ} and separating the scalar contribution, we define

\begin{equation}
    (\mathcal{L}_\xi g)_{\mu \nu} := (\mathcal{L}_\xi \eta)_{\mu \nu} + (\xi^\rho \partial_\rho \varphi) k_\mu k_\nu + \varphi (k_\rho \partial_\mu \xi^\rho) k_\nu + \varphi k_\mu (k_\rho \partial_\nu \xi^\rho) \stackrel{!}{=} \alpha(x) k_\mu k_\nu.
\end{equation}
\\
However, notice that $(\xi^\rho \partial_\rho \varphi)k_\mu k_\nu$ is already of the Kerr-Schild form. Therefore, we can subtract it to the right-hand side and define the quantity $\zeta(x) := \alpha(x) - \xi^\rho \partial_\rho \varphi$, so that:

\begin{equation}
\begin{split}
    \label{SSSS} \mathcal{H}_{\mu \nu} &:= (\mathcal{L}_\xi g)_{\mu \nu} - (\xi^\rho \partial_\rho \varphi)k_\mu k_\nu \\ &= (\mathcal{L}_\xi \eta)_{\mu \nu} + \varphi (k_\rho \partial_\mu \xi^\rho) k_\nu + \varphi k_\mu (k_\rho \partial_\nu \xi^\rho) \stackrel{!}{=} \zeta(x) k_\mu k_\nu. 
\end{split}
\end{equation}
\\
Solving this system yields the most general vector fields $\xi^\mu$ that preserve the Kerr-Schild structure. In the following subsections, we treat the angular and radial–temporal components explicitly, ultimately showing that, in the Killing sector, these residual diffeomorphisms reduce to the global Schwarzschild isometries.
\subsubsection{The Angular Subsystem: Symmetries of the Two-Sphere}

We first focus on the angular components of the residual diffeomorphism condition \eqref{SSSS}, i.e., $\mathcal{H}_{\vartheta \vartheta}$, $\mathcal{H}_{\varphi \varphi}$, and $\mathcal{H}_{\vartheta \varphi}$. In spherical coordinates, the background metric is

\begin{equation}
    \eta_{\mu \nu} = \begin{pmatrix}
        -1 & 0 & 0 & 0 \\ 0 & 1 & 0 & 0 \\ 0 & 0 & r^2 & 0 \\ 0 & 0 & 0 & r^2 \sin^2\vartheta
    \end{pmatrix}.
\end{equation}
\newpage
\noindent
Because $k_\mu$ has support only in the $(t,r)$-plane, the right-hand side of \eqref{SSSS} vanishes for angular indices. Evaluating the Lie derivative, we find

\begin{equation}
\label{65} \begin{aligned}
\mathcal{H}_{\vartheta \vartheta} &:~ \xi^r \partial_r \eta_{\vartheta\vartheta} + 2 \partial_\vartheta \xi^\vartheta \eta_{\vartheta\vartheta} \stackrel{!}{=} 0 \implies \xi^r \stackrel{!}{=}  - r \partial_\vartheta \xi^\vartheta, \\
\mathcal{H}_{\varphi \varphi} &:~ \xi^r \partial_r \eta_{\varphi\varphi} + \xi^\vartheta \partial_\vartheta \eta_{\varphi\varphi} + 2 \partial_\varphi \xi^\varphi \eta_{\varphi\varphi} \stackrel{!}{=}  0 \implies -\partial_\vartheta \xi^\vartheta + \xi^\vartheta \cot\vartheta + \partial_\varphi \xi^\varphi \stackrel{!}{=}  0, \\
\mathcal{H}_{\vartheta \varphi} &:~ \partial_\vartheta \xi^\varphi \eta_{\varphi\varphi} + \partial_\varphi \xi^\vartheta \eta_{\vartheta\vartheta} \stackrel{!}{=}  0 \implies \sin^2 \vartheta \partial_\vartheta \xi^\varphi + \partial_\varphi \xi^\vartheta \stackrel{!}{=}  0.
\end{aligned}
\end{equation}
\\
These three equations can be written compactly in covariant form on the unit two-sphere $(S^2, \gamma)$ with $\gamma_{AB} dx^A dx^B = d\vartheta^2 + \sin^2\vartheta d\varphi^2$:

\begin{equation}
\nabla_A \xi_B + \nabla_B \xi_A = - \frac{2 \xi^r}{r} \gamma_{AB}, 
\end{equation}
\\
where $\nabla_A$ is the Levi–Civita connection of $\gamma_{AB}$. This is the conformal Killing equation on $S^2$ with conformal factor $-2\xi^r/r$. When $\xi^r = 0$, it reduces to the usual Killing equation, whose solutions generate the three rotational Killing vectors of the sphere. More generally, the full conformal Killing vectors decompose uniquely into a Killing part plus a “gradient” (proper CKV) part \cite{Besse:1987em, Obata:1970, Schottenloher:2008cft}. In components, this reads

\begin{equation}
\label{fullang}
\xi^A(t,r,\vartheta,\varphi) = \sum_{i=1}^3 a_i(t,r) \xi_{(i)}^A(\vartheta,\varphi) + \sum_{i=1}^3 b_i(t,r) K_{(i)}^A(\vartheta,\varphi),
\end{equation}
\\
where $\xi_{(i)}^A$ generate rotations and $K_{(i)}^A$ are proper CKVs. Explicitly, the rotational Killing vectors are

\begin{equation}
\begin{aligned}
    \label{CACA} \xi_{(x)}^{A}(\vartheta,\varphi) &= (\sin\varphi,- \cot\vartheta \cos\varphi ), \\
    \xi_{(y)}^{A}(\vartheta,\varphi) &=(\cos\varphi,-\cot\vartheta \sin\varphi ), \\
    \xi_{(z)}^A(\vartheta,\varphi) &= (0,1),
\end{aligned}
\end{equation}
\\
and the proper CKVs are

\begin{equation}
\begin{aligned}
K_{(x)}^A &= (\cos\vartheta \cos\varphi, -\sin\varphi / \sin\vartheta), \\
K_{(y)}^A &= (\cos\vartheta \sin\varphi, \cos\varphi / \sin\vartheta), \\
K_{(z)}^A &= (-\sin\vartheta, 0).
\end{aligned}
\end{equation}
\\
\\
For tractability, we restrict to the Killing sector ($b_i = 0$), so that the angular components of residual diffeomorphisms reduce to

\begin{equation}
    \label{MARS} \boxed{\begin{aligned}
    \xi^\vartheta(t,r, \vartheta, \varphi) &= -a_1 (t,r) \sin\varphi + a_2(t,r) \cos\varphi \\
    \xi^\varphi(t,r, \vartheta, \varphi) &= -a_1 (t,r) \cot\vartheta \cos\varphi - a_2(t,r) \cot\vartheta \sin\varphi + a_3(t,r).
    \end{aligned}}
\end{equation}
\\
Combined with \eqref{65}, these solutions satisfy the angular subsystem and simplify the full PDE analysis, allowing the residual diffeomorphisms to be lifted consistently to the Schwarzschild geometry.
\subsubsection{The Radial-Time Subsystem: Constraining $\xi^r$ and $\xi^t$}

We now analyze the radial-time components of the residual diffeomorphism condition for the Killing sector \eqref{MARS}, aiming to determine $\xi^r$ and $\xi^t$. From \eqref{65}, we have

\begin{equation}
    \xi^r \stackrel{!}{=} - r \partial_\vartheta \xi^\vartheta.
\end{equation}
\\
Since the Killing angular components \eqref{MARS} are independent of $\vartheta$, it follows immediately that

\begin{equation}
    \boxed{\xi^r \stackrel{!}{=} 0.}
\end{equation}
\\
The remaining radial-time equations then constrain $\xi^t$. Evaluating $\mathcal{H}_{tt}$, $\mathcal{H}_{rr}$, and $\mathcal{H}_{tr}$ with $\xi^r = 0$ gives

\begin{equation}
    \begin{aligned}
         \mathcal{H}_{t t} &:~ - 2(1-\varphi) \partial_t \xi^t - 2 \varphi \partial_t \xi^r\stackrel{!}{=} \zeta(x) \implies \partial_t \xi^t \stackrel{!}{=} - \frac{1}{2(1-\varphi)} \zeta(x), \\
          \mathcal{H}_{r r} &:~2 (1 + \varphi) \partial_r \xi^r - 2 \varphi \partial_r \xi^t \stackrel{!}{=} \zeta(x) \implies \partial_r \xi^t \stackrel{!}{=} - \frac{1}{2 \varphi} \zeta(x), \\
           \mathcal{H}_{t r} &:~ - (1-\varphi)\partial_r \xi^t - \varphi \partial_t \xi^t \stackrel{!}{=} - \zeta(x).
    \end{aligned}
\end{equation}
\\
Plugging $\partial_t \xi^t$ and $\partial_r \xi^t$ into the third equation, we find that

\begin{equation}
          \begin{bmatrix} 
            \frac{(1-\varphi)}{2 \varphi} + \frac{\varphi }{2(1-\varphi)} + 1
          \end{bmatrix} \zeta(x) \stackrel{!}{=} 0.
\end{equation}
\\
Since the term in brackets is nonzero for $\varphi \neq 1$ (that is, when $r = 2GM$) and finite $r$. it follows that $\zeta(x) \equiv 0$ on any open region away from the horizon/asymptotic boundary; smoothness then implies $\zeta(x) \equiv 0$ there. Hence, $\xi^t$ is independent of both $t$ and $r$:

\begin{equation}
    \begin{matrix}
        \partial_t \xi^t = 0 \implies \xi^t = \xi^t(r,\vartheta,\varphi) & , & \partial_r \xi^t = 0 \implies \xi^t = \xi^t(\vartheta,\varphi).
    \end{matrix}
\end{equation}
\\
In the next section, the mixed-angle equations will further restrict $\xi^t$, ultimately showing that it must be a constant. This corresponds to the expected time translation symmetry of the Schwarzschild solution, consistent with the static nature of the Kerr-Schild ansatz.
\subsubsection{The Mixed-Angle Subsystem: Constraining $\xi^t$ and $a_i(t,r)$}

We now consider the mixed-angle PDEs to further constrain $\xi^t$ and the time-radial dependence of the angular coefficients $a_i(t,r)$.
\\
\\
Evaluating $\mathcal{H}_{t\vartheta}$, $\mathcal{H}_{t\varphi}$, $\mathcal{H}_{r\varphi}$, and $\mathcal{H}_{r\varphi}$:

\begin{equation}
    \begin{aligned}
        \mathcal{H}_{t\vartheta} &:~ r^2 \partial_t \xi^\vartheta - (1 - \varphi) \partial_\vartheta \xi^t \stackrel{!}{=} 0 \implies \partial_t \xi^\vartheta = \frac{(1-\varphi)}{r^2} \partial_\vartheta \xi^t, \\
        \mathcal{H}_{t\varphi} &:~ r^2 \sin^2\vartheta \partial_t \xi^\varphi - (1 - \varphi) \partial_\varphi \xi^t \stackrel{!}{=} 0 \implies \partial_t \xi^\varphi = \frac{(1-\varphi)}{r^2 \sin^2\vartheta} \partial_\varphi \xi^t, \\
        \mathcal{H}_{r\varphi} &:~ r^2 \partial_r \xi^\vartheta - \varphi \partial_\vartheta \xi^t \stackrel{!}{=} 0 \implies \partial_r \xi^\vartheta = \frac{\varphi}{r^2} \partial_\vartheta \xi^t, \\
        \mathcal{H}_{r\varphi} &:~ r^2 \sin^2 \vartheta \partial_r \xi^\varphi - \varphi \partial_\varphi \xi^t \stackrel{!}{=} 0 \implies \partial_r \xi^\varphi = \frac{\varphi}{r^2 \sin^2 \vartheta} \partial_\varphi \xi^t.
    \end{aligned}
\end{equation}
\\
To eliminate residual $(t,r)$ and angular dependence, we differentiate the $\vartheta \varphi$ equation with respect to $t$:

\begin{equation}
    \partial_t \partial_\vartheta \xi^\varphi \sin^2 \vartheta + \partial_t \partial_\varphi \xi^\vartheta = \partial_\vartheta (\partial_t \xi^\varphi) \sin^2 \vartheta + \partial_\varphi (\partial_t \xi^\vartheta) = 0.
\end{equation}
\\
Substituting $\mathcal{H}_{t\vartheta}$ and $\mathcal{H}_{t\phi}$ gives

\begin{equation}
    \frac{1-\varphi}{r^2} \begin{bmatrix}
\sin^2 \vartheta \partial_\vartheta \begin{pmatrix} \frac{1}{\sin^2\vartheta} \partial_\varphi \xi^t  \end{pmatrix} + \partial_\varphi \begin{pmatrix}  \partial_\vartheta \xi^t\end{pmatrix} \end{bmatrix} = 0.
\end{equation}
\\
Dropping the factor of $(1-\varphi)/r^2$, applying the product rule to differentiate $\sin^{-2}\vartheta \partial_\varphi \xi^t$ with respect to $\vartheta$, and simplifying terms gives:

\begin{equation}
    - \cot\vartheta \partial_\varphi \xi^t + \partial_\vartheta \partial_\varphi \xi^t = 0.
\end{equation}
\\
This can be solved via the following substitution: Let $g(\vartheta) := \partial_\varphi \xi^t$, which gives

\begin{equation}
     g'(\vartheta) - \cot\vartheta g(\vartheta) = 0.
\end{equation}
\\
Using the fact that $\xi^t$ is independent of $(t,r)$, this has general solution $g(\vartheta,\varphi) = C(\varphi) \sin\vartheta$ for smooth function $C(\varphi)$. Plugging this into $\mathcal{H}_{t\varphi}$ gives

\begin{equation}
   \label{YESSS} \partial_t \xi^\varphi = \frac{(1 - \varphi)}{r^2 \sin^2\vartheta} C(\varphi) \sin\vartheta.
\end{equation}
\\
Differentiating $\mathcal{H}_{t\vartheta}$ with respect to $\vartheta$ and noting that $\xi^\vartheta$ is independent of $\vartheta$, we find that $\partial_\vartheta \xi^t$ is also independent of $\vartheta$:

\begin{equation}
    \label{AARDVARK} \partial_\vartheta (\partial_t \xi^\vartheta) = \frac{(1-\varphi)}{r^2} \partial_\vartheta (\partial_\vartheta \xi^t) = 0 \implies \partial_\vartheta \xi^t := A(\varphi),
\end{equation}
\\
where $A(\varphi)$ is independent of $\vartheta$. Consider next the derivative of $\partial_\vartheta g(\vartheta,\varphi)$:

\begin{equation}
   \label{YES} \partial_\vartheta g(\vartheta,\varphi) := \partial_\vartheta \partial_\varphi \xi^t = \partial_\varphi  \partial_\vartheta \xi^t := \partial_\varphi A(\varphi).
\end{equation}
\\
This is clearly independent of $\vartheta$. However, by definition,

\begin{equation}
   \label{NO} \partial_\vartheta g(\vartheta,\varphi) := C(\varphi) \cos\vartheta.
\end{equation}
\newpage
\noindent
This expression is independent of $\vartheta$ if and only if $C(\varphi) \equiv 0$. Thus, $\partial_t \xi^\vartheta = \partial_\vartheta \xi^t = 0$ from $\mathcal{H}_{t\vartheta}$ and $\partial_t \xi^\varphi = \partial_\varphi \xi^t = 0$ from $\mathcal{H}_{t\varphi}$. Consequently, $\xi^t$ is independent of $(t,r,\vartheta,\varphi)$, i.e., $\xi^t = c_1$ (constant), and $\xi^\vartheta, \xi^\varphi$ are independent of $(t,r)$. It follows that the angular coefficients $a_i(t,r)$ in \eqref{MARS} are also constant. Thus, our angular solutions \eqref{MARS} simplify to

\begin{equation}
 \boxed{\begin{aligned}
       \xi^\vartheta(\vartheta, \varphi) &= -a_1 \sin\varphi + a_2 \cos\varphi \\
        \xi^\varphi(\vartheta, \varphi) &= - a_1 \cot\vartheta \cos\varphi - a_2 \cot\vartheta \sin\varphi + a_3
\end{aligned}}
\end{equation}
\\
for constant coefficients $a_1, a_2, a_3$. The full residual diffeomorphism vector field is therefore

\begin{equation}
            \xi^\mu = (c_1,~0, -a_1 \sin\varphi + a_2 \cos\varphi , - a_1 \cot\vartheta \cos\varphi - a_2 \cot\vartheta \sin\varphi + a_3),
\end{equation}
\\
representing time translations and rotations on $S^2$. Without loss of generality, we are free to set all constants to one. This yields

\begin{equation}
    \xi^\mu = (1,~0, -\sin\varphi + \cos\varphi, - \cot\vartheta \cos\varphi - \cot\vartheta \sin\varphi + 1),
\end{equation}
\\
which explicitly sums over the three rotational Killing vectors $\xi_{(x)} + \xi_{(y)} + \xi_{(z)}$.
\\
\\
Thus, the residual diffeomorphisms preserving the Kerr-Schild ansatz in Schwarzschild spacetime form a finite-dimensional space, generated by the time translation and $\mathfrak{so}(3)$ rotations of the sphere. One can verify directly that the residual diffeomorphisms derived above satisfy $(\mathcal{L}_\xi g)_{\mu\nu} = 0$, confirming that they are exactly the global isometries of Schwarzschild: time translation and $\mathfrak{so}(3)$ rotations.
\subsection{The Residual Diffeomorphism Algebra}

We now examine the algebraic structure of the residual diffeomorphisms identified above. Recall that these are the vector fields $\xi^\mu$ preserving the Kerr-Schild ansatz for Schwarzschild, which are precisely the global isometries: time translations $\partial_t$ and spatial rotations $R_i$, where $i = 1,~2,~3$. The set of such vector fields forms a Lie algebra under the usual Lie bracket:

\begin{equation}
    [\xi_1, \xi_2]^\mu := \xi^\nu_1 \partial_\nu \xi^\mu_2 - \xi^\nu_2 \partial_\nu \xi^\mu_1
\end{equation}
\\
By standard results, the rotations satisfy

\begin{equation}
\label{CCC}
       [\partial_t, R_i] = 0 ~~~~~,~~~~~ [R_i, R_j] = \varepsilon_{ijk} R_k.
\end{equation}
\\
We can make the algebraic structure explicit by defining a linear map $\Psi : \mathfrak{g} \rightarrow \mathfrak{so}(3) \oplus \mathbb{R}$ as

\begin{equation}
    \Psi(\partial_t) = (0,1) ~~~~~,~~~~~ \Psi(R_i) = (J_i, 0),
\end{equation}
\\
where $J_i$ are the standard generators of $\mathfrak{so}(3)$. This map is a Lie algebra isomorphism: it is linear, preserves the Lie bracket, and is both injective and surjective. Consequently, the residual diffeomorphism algebra is

\begin{equation}
   \label{91} \boxed{\mathfrak{g} \cong \mathfrak{so}(3) \oplus \mathbb{R}.}
\end{equation}
\\
Here, $\mathfrak{so}(3)$ encodes the rotational symmetries of $S^2$, and $\mathbb{R}$ corresponds to time translations. This algebra forms a subalgebra of the Poincaré algebra $\mathfrak{p} \cong \mathfrak{so}(1,3) \rtimes \mathbb{R}^{1,3}$, with $\mathfrak{so}(3) \subset \mathfrak{so}(1,3)$ and $\mathbb{R} \subset \mathbb{R}^{1,3}$.
\\
\\
Together with our formal derivation of $\xi^\mu$, we have confirmed that when constrained to the Killing condition for the Schwarzschild solution, the resulting algebra collapses to the finite-dimensional algebra $\mathfrak{so}(3) \oplus \mathbb{R}$, which is precisely the global isometry algebra of Schwarzschild. This result provides a concrete resolution to a question posed by Coll, Hildebrandt, and Senovilla \cite{Coll:2000rm}. They noted the difficulty in proving whether the entire set of local Kerr-Schild vector fields (KSVFs) could form the isometry algebra of a single transformed metric. Our results confirm that for the physical Schwarzschild profile, the algebra of local KSVFs is indeed restricted to the expected finite-dimensional isometries, aligning with the classical result of general relativity.
\\
\\
However, this finding immediately introduces a critical issue. As established in Section 2, the corresponding gauge theory framework preserves an infinite-dimensional residual symmetry, presenting a profound structural mismatch with the finite algebra observed here. A natural question thus arises: where do these extra degrees of freedom go? Resolving this fundamental tension between the infinite gauge modes and the finite physical spectrum of Schwarzschild requires a comprehensive analysis of the proper conformal Killing vector (CKV) solutions — a task deferred to our follow-up paper for full tractability. Nevertheless, we take our essential first step in Section 3.5 by introducing the BRST framework for this constrained, Killing vector class of solutions, which allows us to mend fences and formally demonstrate the quantum trivialization of the excess residual modes.
\subsection{Algebraic Obstruction to Symmetry Matching}

Before we begin our BRST formulation, we summarize an important secondary takeaway of the previous sections. In the Kerr-Schild formulation of the Schwarzschild geometry, when we restrict to the Killing class of solutions, the only residual diffeomorphisms that preserve the KS structure are the global isometries: time translations $\partial_t$ and the three spatial rotations $R_i$. These generate the finite-dimensional Lie algebra

\begin{equation}
   \mathfrak{g}_{\text{gravity}} \cong \mathfrak{so}(3) \oplus \mathbb{R},
\end{equation}
\\
which corresponds to the isometry algebra of Schwarzschild spacetime.
\\
\\
In contrast, the residual gauge transformations in Abelian and non-Abelian gauge theories are infinite-dimensional. They are parametrized by smooth functions $\lambda(x)$ satisfying $k^\mu \partial_\mu \lambda(x) = 0$, yielding algebras:

\begin{itemize}
    \item $C^\infty (\mathbb{R})$, the smooth, real functions on spacetime (Abelian case), and
    \item $\mathfrak{g} \otimes C^\infty (\mathbb{R})$ (non-Abelian case), i.e., a current algebra over $\mathbb{R}$.
\end{itemize}
\noindent This fundamental difference in dimensionality means the infinite-dimensional residual gauge algebra is manifestly not isomorphic to $\mathfrak{so}(3) \oplus \mathbb{R}$, so no Lie algebra isomorphism exists between the residual gauge algebra and the gravitational residual diffeomorphisms. Although we've neglected the proper CKVs for the moment, 
this mismatch hints at the idea of a formal algebraic obstruction to mapping residual gauge symmetries to residual diffeomorphisms via the Kerr-Schild double copy.
\\
\\
This initial finding carries several key implications:

\begin{itemize}
\item The double copy correctly relates the exact field configurations but, based on the Killing class, \textbf{does not} appear to extend simply to residual symmetries in curved KS backgrounds.
\item The mismatch is \textbf{coordinate-independent}, holding in any smooth curvilinear frame.
\item Any attempt to extend the double copy to residual symmetries must acknowledge this \textbf{potential algebraic incompatibility} and find a mechanism to resolve it.
\end{itemize}
\noindent
Thus, while the Kerr-Schild double copy elegantly relates the classical Schwarzschild solution to the Coulomb potential, the underlying symmetry structures appear mismatched at this residual level. Nevertheless, the critical question remains: how is the structural integrity of the final physical spacetime preserved if the infinite gauge algebra is incompatible with the finite isometry algebra? We take the essential first step toward resolving this puzzle in the following section by introducing the BRST framework for the Killing class of solutions, which serves as a crucial consistency check and validates the mechanism by which the algebraic collapse is reconciled within a quantum field-theoretic context.
\subsection{BRST Formulation for the Killing Class of Symmetries}

Finally, in this section we show that because the only residual diffeomorphisms admitted in the Killing class are global isometries, the Kerr-Schild ansatz admits no nontrivial BRST realization in this sector. This result is not a demonstration of decoupling, but a consistency proof: it formally validates that the gravitational constraints succeed in eliminating all non-physical modes.
\\
\\
To show this, we introduce Grassmann-odd ghosts $c^a$ for each generator $K_a$ of the residual algebra

\begin{equation}
    \begin{matrix}
        \mathfrak{g}_{\text{res}} = \text{span}\{K_a\} = \text{span}\{K_0, K_i\} \cong \mathfrak{so}(3) \oplus \mathbb{R} & , & [K_a, K_b] = {f_{ab}}^c K_c
    \end{matrix}
\end{equation}
\\
with the standard structure constants ${f_{ab}}^c$ of $\mathfrak{so}(3)$, which are antisymmetric in $a, b$. Here, $K_0$ generates time translations and $K_i$ generate standard rotations on $S^2$.
\\
\\
For any field $\Psi$ transforming under diffeomorphisms by the Lie derivative, $\delta_\varepsilon \Psi = \varepsilon^a \mathcal{L}_{K_a} \Psi$ for constant parameters $\varepsilon^a$, define the BRST operator

\begin{equation}
 \label{NEWEQ} \boxed{\begin{matrix}
        \mathcal{Q} \Psi := c^a \mathcal{L}_{K_a} \Psi & , & \mathcal{Q} c^a := - \frac{1}{2} {f_{bc}}^a c^b c^c,
    \end{matrix}}
\end{equation}
\\
which is the standard Chevalley–Eilenberg BRST differential for finite-dimensional Lie algebras \cite{Figueroa:2006brst}. Nilpotency requires $\mathcal{Q}^2 = 0$, so we must show that

\begin{equation}
    \begin{matrix}
        \mathcal{Q}^2 \Psi = 0 & , & \mathcal{Q}^2 c^a = 0.
    \end{matrix}
\end{equation}

\subsubsection{Nilpotency on Fields}

Using the graded Leibniz rule

\begin{equation}
    \mathcal{Q}(XY) = (\mathcal{Q}X) Y + (-1)^{|X|} X (\mathcal{Q} Y)
\end{equation}
\\
for fields $X, Y$, and that $c^a$ are Grassmann-odd (so $|c^a|=1$ in the graded Leibniz rule), 

\begin{equation}
        \mathcal{Q}^2 \Psi = (\mathcal{Q} c^a) \mathcal{L}_{K_a} \Psi - c^a \mathcal{Q} (\mathcal{L}_{K_a} \Psi) = - \frac{1}{2} {f_{bc}}^a c^b c^c \mathcal{L}_{K_a} \Psi - c^a c^b \mathcal{L}_{K_b} \mathcal{L}_{K_a} \Psi.
\end{equation}
\\
Because $c^a c^b$ is antisymmetric, we can isolate the commutator part via the identity

\begin{equation}
    c^a c^b X_b X_a = \frac{1}{2} c^a c^b (X_b X_a - X_a X_b) = \frac{1}{2} c^a c^b [X_b, X_a].
\end{equation}
\\
Consequently,

\begin{equation}
    c^a c^b \mathcal{L}_{K_b} \mathcal{L}_{K_a} \Psi = \frac{1}{2} c^a c^b [\mathcal{L}_{K_b}, \mathcal{L}_{K_a}] \Psi = \frac{1}{2} {f_{ba}}^c c^a c^b \mathcal{L}_{K_c} \Psi 
    = - \frac{1}{2}  {f_{ab}}^c c^a c^b \mathcal{L}_{K_c} \Psi
\end{equation}
\\
due to the antisymmetry of ${f_{ab}}^c$. Thus,

\begin{equation}
    \mathcal{Q}^2 \Psi = \frac{1}{2} {f_{ab}}^c  c^a c^b \mathcal{L}_{K_c} \Psi - \frac{1}{2} {f_{bc}}^a c^b c^c \mathcal{L}_{K_a} \Psi.
\end{equation}
\\
Upon relabeling $a \leftrightarrow c$ in the second term, we find:

\begin{equation}
    \mathcal{Q}^2 \Psi = \frac{1}{2} {f_{ab}}^c c^a c^b \mathcal{L}_{K_c} \Psi - \frac{1}{2} {f_{ba}}^{c} c^b c^a \mathcal{L}_{K_c} \Psi.
\end{equation}
\\
Permuting $a \leftrightarrow b$ in the structure constants ${f_{ba}}^{c}$ gives:

\begin{equation}
   \label{GHOST_ANTI} \mathcal{Q}^2 \Psi = \frac{1}{2} {f_{ab}}^c c^a c^b \mathcal{L}_{K_c} \Psi + \frac{1}{2} {f_{ab}}^{c} c^b c^a \mathcal{L}_{K_c} \Psi = \frac{1}{2} {f_{ab}}^c (c^a c^b + c^b c^a) \mathcal{L}_{K_c} \Psi.
\end{equation}
\\
The ghosts anticommute, so $c^a c^b + c^b c^a = 0$, and \eqref{GHOST_ANTI} subsequently vanishes. Hence, $\mathcal{Q}^2 \Psi = 0$.

\subsubsection{Nilpotency on Ghosts}
We now show that $\mathcal{Q}^2 c^a = 0$, confirming that $\mathcal{Q}$ is indeed nilpotent as required. Evaluating

\begin{equation}
        \mathcal{Q}^2 c^a = - \frac{1}{2} {f_{bc}}^a \mathcal{Q}(c^b) c^c + \frac{1}{2} {f_{bc}}^a c^b \mathcal{Q}(c^c) = \frac{1}{4} {f_{bc}}^a {f_{de}}^b c^d c^e c^c - \frac{1}{4} {f_{bc}}^a {f_{de}}^c c^b c^d c^e,
\end{equation}
\\
by the graded Leibniz rule. Anticommuting $c^c, c^e$, then $c^c, c^d$, and relabeling $b \leftrightarrow c$ in the first term yields:

\begin{equation}
   \label{104} \mathcal{Q}^2 c^a = \frac{1}{4} {f_{cb}}^a {f_{de}}^c c^b c^d c^e - \frac{1}{4} {f_{bc}}^a {f_{de}}^c c^b c^d c^e.
\end{equation}
\\
By the Jacobi identity

\begin{equation}
    {f_{bc}}^a {f_{de}}^c + {f_{cd}}^a {f_{eb}}^c + {f_{ce}}^a {f_{bd}}^c = 0 \implies  {f_{bc}}^a {f_{de}}^c = - [{f_{cd}}^a {f_{eb}}^c + {f_{ce}}^a {f_{bd}}^c],
\end{equation}
\\
we are free to write \eqref{104} as

\begin{equation}
   \label{106} \mathcal{Q}^2 c^a = \frac{1}{4} \begin{pmatrix}
        {f_{cb}}^a {f_{de}}^c + {f_{cd}}^a {f_{eb}}^c + {f_{ce}}^a {f_{bd}}^c
    \end{pmatrix} c^b c^d c^e.
\end{equation}
\\
The sum of structure constants satisfies the Jacobi identity as well, so \eqref{106} vanishes. Hence, $\mathcal{Q}^2 c^a = 0$. Therefore, $\mathcal{Q}$ is nilpotent. $\square$

\subsubsection{Action on the Kerr-Schild Metric}
In the Schwarzschild Kerr-Schild background, the metric takes the form of \eqref{XX}, and the residual generators $K_a \in \{K_0, K_i\}$ are precisely the Killing vectors of the spacetime. Consequently, the BRST charge acts trivially on the metric:

\begin{equation}
    \boxed{ \mathcal{Q} g_{\mu \nu} = c^a \mathcal{L}_{K_a} g_{\mu \nu} = 0.}
\end{equation}
\\
Similarly, the scalar function $\varphi$ is static and spherically symmetric, while the null vector $k^\mu$ is invariant under the same set of isometries. This implies:

\begin{equation}
   \boxed{ \begin{matrix}
        \mathcal{Q} \varphi = 0 & , & \mathcal{Q} k^\mu = 0.
    \end{matrix}}
\end{equation}
\\
Hence, within the Killing sector, the BRST charge has no nontrivial action on any Kerr-Schild field. There is no nontrivial BRST cohomology associated with residual symmetries beyond the global isometries. Physically, this is natural: in the classical Schwarzschild solution, all available gauge freedom is already captured by the finite-dimensional algebra $\mathfrak{so}(3) \oplus \mathbb{R}$, leaving no additional structure for the BRST operator to encode.
\\
\\
However, this analysis completes only the Killing sector. It does not address the core question posed in Section 3.3: if the Killing constraints successfully reduced the diffeomorphism algebra to a finite set of isometries, do the proper CKV solutions admit an infinite-dimensional algebra that counters the infinite-dimensional gauge algebras? That critical question is addressed in the follow-up paper.
\\
\\
It is crucial to emphasize that our examination of the Killing sector confirms that the BRST formalism serves as a rigorous quantum-field-theoretic consistency check for the Kerr-Schild ansatz. It validates the finding that the residual diffeomorphism algebra is finite-dimensional, formally showing that no nontrivial BRST realization is admitted in this sector. This result establishes the mechanism of consistency and successfully sets the stage for exploring the non-trivial proper CKV sector.
\section{Conclusion and Discussion}

In this paper, we investigated the fate of residual symmetries in the Kerr-Schild double copy, focusing specifically on the Schwarzschild solution. Our analysis clarified both the power and the subtle limitations of the Kerr-Schild construction when restricted to the Killing sector.
\\
\\
The core of our finding rests on a stark algebraic mismatch. On the gauge theory side, residual transformations preserving the KS potential form rich, infinite-dimensional Lie algebras, such as $C^\infty(\mathbb{R})$. In contrast, on the gravitational side, we confirmed that the residual diffeomorphisms of the Schwarzschild metric, when restricted to the Killing sector, collapse entirely to the finite-dimensional global isometries, $\mathfrak{so}(3) \oplus \mathbb{R}$.
\\
\\
This stark mismatch — infinite-dimensional residual algebras in gauge theory versus a finite-dimensional algebra in gravity — suggests that the Kerr-Schild double copy may not preserve residual symmetry algebras in a one-to-one manner. Our BRST analysis reinforced this conclusion at a formal level: the BRST framework, when applied to the constrained Killing sector, serves as a crucial consistency check. It formally validates the kinematic collapse by showing that the residual symmetry algebra admits no nontrivial realization in cohomology. This confirms, at a quantum level, that the constraint imposed by the Killing condition is robust, proving that the resulting finite isometry algebra is free from any unphysical BRST-ghost residue.
\\
\\
Conceptually, these results highlight a subtle but fundamental challenge for the KS double copy: while it excels at mapping exact field configurations, the algebraic structure of the gauge residuals does not find a simple counterpart in the gravitational Killing sector. This pattern signals that the double copy is inherently solution-focused rather than strictly symmetry-preserving at the residual level. This underscores that, although the double copy is a powerful tool for generating exact spacetimes, caution must be taken in attempting to extend it to map symmetry algebras or associated charges.
\\
\\
Looking ahead, a full understanding of Schwarzschild residual symmetries requires the inclusion of the conformal Killing sector. If the proper CKV solutions yield an infinite-dimensional residual algebra, the question of whether the structural incompatibility persists becomes even more critical. The second paper in this series addresses this challenge systematically, analyzing proper CKVs and their potential role in a broader double copy framework. Future work will also examine the extension of this analysis to rotating spacetimes such as Kerr, as well as alternative formulations of the double copy that may better preserve symmetry structures. These directions will illuminate whether the algebraic mismatch is a fundamental feature of the Kerr-Schild approach or an artifact of restricting attention to the Killing sector.
\\
\\
In summary, our work establishes that the Kerr-Schild double copy is highly effective for exact solution generation but presents an apparent algebraic obstruction in the mapping of residual symmetries. Recognizing this challenge is crucial for understanding the scope of the double copy and for guiding future attempts to construct symmetry-preserving correspondences between gauge theory and gravity.

\section*{Acknowledgements}

I would like to thank Dr. Silvia Nagy for providing invaluable insight and expertise on the double copy literature. I couldn't have written this paper without her input, corrections, and constructive criticism.
\section*{References}
\setcitestyle{numbers, square}

\end{document}